# First-principles study on the effective masses of zinc-blend-derived $Cu_2Zn-IV-VI_4$ (IV = Sn, Ge, Si and VI = S, Se)

Heng-Rui Liu,[1] Shiyou Chen,[2] Ying-Teng Zhai,[1] H. J. Xiang,[1] X. G. Gong[1] and Su-Huai Wei[3]

1. Key Laboratory for Computational Physical Sciences (MOE), State Key Laboratory of Surface Physics, and Department of Physics, Fudan University, Shanghai 200433, China
2. Key Laboratory of Polar Materials and Devices (MOE), East China Normal University, Shanghai 200241, China
3. National Renewable Energy Laboratory, Golden, Colorado 80401, USA

## Abstract

The electron and hole effective masses of kesterite (KS) and stannite (ST) structured $Cu_2Zn-IV-VI_4$ (IV = Sn, Ge, Si and VI = S, Se) semiconductors are systematically studied using first-principles calculations. We find that the electron effective masses are almost isotropic, while strong anisotropy is observed for the hole effective mass. The electron effective masses are typically much smaller than the hole effective masses for all studied compounds. The ordering of the topmost three valence bands and the corresponding hole effective masses of the KS and ST structures are different due to the different sign of the crystal-field splitting. The electron and hole effective masses of Se-based compounds are significantly smaller compared to the corresponding S-based compounds. They also decrease as the atomic number of the group IV elements (Si, Ge, Sn) increases, but the decrease is less notable than that caused by the substitution of S by Se.

The quaternary semiconductors $Cu_2Zn-IV-VI_4$ (IV = Si, Ge, Sn and VI = S, Se) have received wide attention as potential solar cell absorber materials. The reported band gaps of these materials are around 1.0~3.0 eV,[1-4] and the $Cu_2ZnSn(S,Se)_4$ solar cells have been fabricated with an efficiency factor above ~10%,[5] which make them as possible substitutes of the currently commercialized $Cu(In,Ga)Se_2$ (CIGSe) solar cells, which may be limited by the scarcity of In.

An ideal solar energy absorber should have a band gap around 1.5 eV and a high optical absorption coefficient for the visible light. With this in mind, the band gaps and optical properties of the $Cu_2Zn-IV-VI_4$ systems have been extensively studied both experimentally[1-14] and theoretically.[15-22] However, the performance of a solar cell absorber does not just depend on its band gap and optical adsorption but also on the electric transport efficiency of carriers (electron and hole), which depends on the effective mass. So far, there are only a few first-principles study[15] on the effective masses of $Cu_2ZnSnS_4$ (CZTS) and $Cu_2ZnSnSe_4$ (CZTSe) and neither experimental nor theoretical results are available for the rest of these quaternary semiconductors.

In this paper, we present a systematic first-principles study on the effective masses of the zinc-blend-derived kesterite (KS) and stannite (ST) structures of $Cu_2Zn-IV-VI_4$ within the density functional formalism as implemented in the VASP code.[23] For the exchange-correlation potential, we used the standard Heyd-Scuseria-Ernzerhof (HSE06) hybrid functional[24] with the screening parameter set to 0.2 Å. The projector augmented-wave (PAW) pseudopotentials[25] with an energy cutoff of 400 eV for the planewave basis and 5×5×5 Monkhorst-Pack[26] Γ-centered k-point meshes were employed to give converged results. The Sn 4d and Ge 3d electrons were treated as valence electrons and the spin-orbit coupling (SOC) was included in the calculations of electronic structures.

Theoretically, the zinc-blend-derived kesterite (KS) structures [space group $I\bar{4}$, Figure 1(a)] have been shown to be the most stable structures for the $Cu_2Zn-IV-VI_4$ systems[16] due to the small strain energy and more negative Madelung energy.[17-20] The zinc-blend-derived stannite (ST) structures [space group $I\bar{4}2m$, Figure 1(b)] are slightly less stable than the kesterite (KS) structures, and the KS and ST ordering may coexist in the synthesized samples.[14]

The calculated energies, lattice constants, band gaps, crystal-field and spin-orbit splittings[27] for zinc-blend-derived KS and ST structures of $Cu_2Zn-IV-VI_4$ (IV = Sn, Ge, Si and VI = S, Se) along with available experimental data are summarized in Table I. Our results show that the KS structures are energetically more favorable than the ST structures, which is consistent with previous findings.[16,18,21] The lattice constants produced by the HSE06 functional are in good agreement with experiments and deviations are typically less than 1%. The tetragonal distortion parameter, $\eta = c/2a$, is less than 1 for the KS structure while lager than 1 for the ST structure due to the different arrangement of Cu and Zn atoms in these two structures and the different bond lengths of the Cu-X, Zn-X, and IV-X bonds.

The calculated band gaps of $Cu_2Zn-IV-VI_4$ using HSE06 also agree well with available experimental findings. The ground state KS structures have the largest gap for all studied components. Although the Se based compounds have notably smaller band gap than the S based compounds, the band gap differences between the KS and the ST structure are almost the same for $Cu_2Zn-IV-S_4$ and $Cu_2Zn-IV-Se_4$ (IV = Sn, Ge). The value is 0.18 eV for $Cu_2ZnSn(S,Se)_4$ and 0.30 eV for $Cu_2ZnGe(S,Se)_4$, respectively. The density of states (DOS) of the $Cu_2ZnGeSe_4$ (CZGSe) is plotted in Figure 2. The shapes of the DOS for the rest quaternary semiconductors $Cu_2Zn-IV-VI_4$ are qualitatively the same (not shown) due to their similar chemical compositions and structures. From the calculated DOS, it can be clearly seen that the DOS of the KS and ST structure are quite similar, the upper valence bands are mainly derived from the hybridization between p states from Se atoms and d states from Cu atoms, while the lower conduction bands are mainly derived from the hybridization between the s and p states from Se atoms and s states from Ge atoms, which are consistent with previous works[15,18,20,21] on $Cu_2ZnSnS_4$ (CZTS) and $Cu_2ZnSnSe_4$ (CZTSe).

Similar to the situation of $Cu(In,Ga)(S,Se)_2$,[28] for Se-based compounds, the spin-orbit splitting are positive and large, while negative and much smaller spin-orbit splittings are observed in S-based materials, which is consistent with the fact that selenium is much heavier than sulphur so relativistic effects are more pronounced in Se-based compounds and that more Cu d components exist in the sulphides. Moreover, the sign of the crystal-field splitting is different between the KS and ST structured $Cu_2Zn-IV-VI_4$, which leads to a different character of the hole effective masses of the topmost 3 valence bands in those structures (see below).

The valence band maximum (VBM) and the conductor band minimum (CBM) of the studied $Cu_2Zn-IV-VI_4$ are all at the $\Gamma$ point. We can write the effective mass tensor of zinc-blend-derived KS and ST structures as the following form by considering the point group symmetry of the crystal cell:

$$\frac{1}{m^*} = \frac{1}{\hbar^2}\begin{pmatrix} a & 0 & 0 \\ 0 & a & 0 \\ 0 & 0 & c \end{pmatrix}.$$

That is, the tensor contains only two independent components: the longitudinal mass $m^{\parallel}$ parallel to the c-axis of the crystal cell shown in Figure 1 and the transverse mass $m^{\perp}$ perpendicular to the c-axis, respectively. One only needs to do one-dimensional fittings along certain directions to get the whole effective mass tensors of these structures.

The results on the electron and hole effective masses at the $\Gamma$ point are summarized in Table II. Our results on the effective masses of $Cu_2ZnSn-VI_4$ (VI = S, Se) are close to Persson's previous findings.[15]

It can be clearly seen from Table II that the electron effective masses are fairly isotropic, while the hole effective masses show strong anisotropy, and the electron effective masses are typically much smaller than the hole effective masses for all studied quaternary semiconductors. Take the effective masses of the KS CZGS (CZTS) as examples: the electron effective masses are $m_c^{\parallel} = 0.21m_e(0.18m_e)$ and $m_c^{\perp} = 0.22m_e(0.19m_e)$, and the hole effective masses for the first VB are $m_{v1}^{\parallel} = 0.27m_e(0.22m_e)$ and $m_{v1}^{\perp} = 0.91m_e(0.74m_e)$, i.e., the transverse hole masses are about three times larger than that of the longitudinal hole mass in these materials. This difference between the electron and the hole effective masses could be understood by analyzing the constituents of the CBM and the VBM shown in Figure 2. For the CBM, a notable contribution from the spherically symmetric s states of the group IV element is observed, which could account for the small isotropic electron effective masses. On the other hand, the VBM is dominated by the low lying anisotropic d states of Cu atoms and p states of the anions, which could lead to the large anisotropic hole masses.

In Figure 3, we plot the topmost 3 VBs along the c-axis of the crystal cell in the immediate neighborhood of the $\Gamma$ point for the KS and the ST CZGS, respectively. The results of other compounds are similar. It can be seen from Figure 3 (a) that for the KS structure, the 1st VB has a much stronger upward energy dispersion at the $\Gamma$ point than those of the next 2 VBs, which results in a much smaller hole effective mass along the c-axis ($m_{v1}^{\parallel} = 0.27 m_e$) of the 1st VB than the next 2 VBs. Moreover, the hole masses along the c-axis of the next 2 VBs are very close to each other: $m_{v2}^{\parallel} = 0.65 m_e \approx m_{v3}^{\parallel} = 0.63 m_e$, since they are split from the doubly-degenerate $\Gamma_{5v}$ states by the spin-orbit coupling (SOC), which is quite weak in S-based compounds. In contrast to the KS CZGS, the ST CZGS owns a opposite ordering in the bands and thus the hole effective masses (see Figure 3 (b)), that is, the hole masses along the c-axis of the topmost 2 VBs ($m_{v1}^{\parallel} = 0.67 m_e \approx m_{v2}^{\parallel} = 0.63 m_e$) are much larger than that of the 3rd VB ($m_{v3}^{\parallel} = 0.25 m_e$).

The differences in the ordering of the topmost VBs and the resulted hole effective masses between the KS and ST structure are caused by the different sign of the crystal-field splitting,[20,27] $\Delta_{cf}$, in those two structures (see Table I). In order to illustrate it, the partial charge density at the $\Gamma$ point of the topmost 3 VBs in the Cu-S-Zn plane for the KS and the Cu-S plane for the ST CZGS are plotted in Figure 4. One can see that for the KS structure, the p orbitals in sulfur of the 1st VB are parallel to the c-axis of the unit cell and those of the next 2 VBs are perpendicular to the c-axis, while for the ST structure, the p orbitals in sulfur of the topmost 2 VBs are perpendicular to the c-axis and those of the 3rd VB are parallel to the c-axis, which clearly indicates that the 1st VB of the KS structure corresponds to the 3rd VB of the ST structure, and the 2nd and the 3rd VB of the KS structure corresponds to the 1st and the 2nd VB of the ST structure, respectively.

The most obvious trend on the effective masses of $Cu_2Zn - IV - VI_4$ (IV = Sn, Ge, Si and VI = S, Se) from Table II is that the electron and hole effective masses of Se-based compounds are significantly smaller than those of S-based compounds, more precisely, the effective masses of $Cu_2Zn - IV - S_4$ are around 50%~100% larger than those of $Cu_2Zn - IV - Se_4$ with the same structure. This trend on the effective masses reflects the fact that the lowest conduction band (CB) and topmost valence band (VB) of Se-based compounds have stronger downward and upward energy dispersions around the $\Gamma$ point than those of S-based compounds, which also contributes to the smaller gaps in Table I. To see it more clearly, the situation for the $Cu_2ZnGe(S,Se)_4$ is plotted in Figure 5, if we look at the effective mass $m_{v1}$ of the first valence band (red line in Figure 5) for the KS structure, the component $m_{v1}^{\perp}$ of the hole effective mass tensor is 0.91 $m_e$ for the CZGS and is only 0.38 $m_e$ for the CZGSe, so the holes in the CZGSe could response to an electrical field applied perpendicular to the c-axis twice faster than those in the CZGS according to the semi-classical model of carriers in crystal.

The above trends can be explained according to two factors: (i) the s and p states of the group VI anions contribute dominantly to the lowest CBs and topmost VBs in $Cu_2Zn - IV - VI_4$, and the s and p states are much more localized in S atom than those in Se atom due to the smaller atomic radius and the resulting stronger attraction by nucleus, (ii) the lattice constants for S-based compounds are around 5% smaller than the Se-based compounds, which means larger band gap and less coupling between valence band and conduction band in S-based compounds. As a result, the electrons and holes in $Cu_2Zn - IV - S_4$ are bound more strongly than those in $Cu_2Zn - IV - Se_4$, which finally leads to lower mobility and larger effective masses in $Cu_2Zn - IV - S_4$. Previously a similar trend has been found by Persson on the effective masses of CIGS and CIGSe.[28]

The effective masses also increase with the decrease of the atomic number of the group IV cation in $Cu_2Zn - IV - VI_4$ (IV = Sn, Ge, Si and VI = S, Se) in most cases, which is illustrated in

Figure 6 for the Se-based compounds, and the results for S-based compounds are qualitatively the same. For example, the hole effective masses perpendicular to the c-axis of the KS $Cu_2Zn-IV-Se_4$ is 0.32, 0.38 and 0.58 $m_e$ for IV = Sn, Ge and Si, respectively. Since the s and p states in the IV cation contribute to the topmost VBs and lowest CBs, respectively, the trend can also be explained by the above arguments.

One can see from Table II that the influences on effective masses by substituting the IV cations are typically smaller than the influences by substituting the group VI anions, for example, the $m_{c1}^{\perp}$ of the KS CZSiS (0.24 $m_e$) is only 26% larger than that of the KS CZTS (0.19 $m_e$) while 60% larger than that of the KS CZSiSe (0.15 $m_e$), this is because that the contribution to the lowest CBs and topmost VBs by the group IV cations are less notable than those of the VI anions, which is clearly illustrated in Figure 2. One can also tell from Table II and more straightforwardly from the slopes in Figure 6 that difference between the effective masses of the $Cu_2ZnSi-VI_4$ and the $Cu_2ZnGe-VI_4$ is larger than that between the $Cu_2ZnGe-VI_4$ and the $Cu_2ZnSn-VI_4$, and the latter is negligible in most cases. As a result, the $Cu_2ZnSnSe_4$ (CZTSe) and the $Cu_2ZnGeSe_4$ (CZGSe) have the smallest effective masses and thus the best mobility of electrons and holes among all studied quaternary semiconductors. Note that the effective masses of the $Cu_2ZnSnSe_4$ (CZTSe) and $Cu_2ZnGeSe_4$ (CZGSe) are also very close to those of the CIGSe.[28]

In conclusion, the effective masses of $Cu_2Zn-IV-VI_4$ (IV = Sn, Ge, Si and VI = S, Se) were studied using density functional methods. We find that (i) for all studied compounds, the electron effective masses are fairly isotropic, while the hole effective masses show strong anisotropy, and the electron effective masses are typically much smaller than the hole effective masses, (ii) the ordering of the hole effective masses of the topmost 3 VBs for the KS and the ST structure are different due to the different sign of the crystal-field splitting in those structures, (iii) the effective masses of Se-based compounds are significantly smaller than the corresponding S-based compounds, substituting the small IV element with larger ones can also decrease the effective masses but the influences are smaller, (iv) $Cu_2ZnSnSe_4$ (CZTSe) and $Cu_2ZnGeSe_4$ (CZGSe) are most suitable solar materials among all studied compounds according to their transport properties of the electrical carriers.

This work is partially supported by the Special Funds for Major State Basic Research, National Science Foundation of China, Ministry of Education and Shanghai Municipality. The calculations were performed in the Supercomputer Center of Fudan University. The work at NREL was funded by the U.S. Department of Energy under the Grant No. DE-AC36-08GO28308.

# References

1. H. Matsushita, T. Ichikawa, and A. Katsui, J. Mater. Sci. 40, 2003 (2005).
2. SeJin Ahn, Sunghun Jung, Jihye Gwak, Ara Cho, Keeshik Shin, Kyunghoon Yoon, Doyoung Park, Hyeonsik Cheong, and Jae Ho Yun, Appl. Phys. Lett. 97, 021905 (2010).
3. Grayson M. Ford, Qijie Guo, Rakesh Agrawal, and Hugh W. Hillhouse, Chem. Mater. 23, 2626 (2011).
4. O. V. Parasyuk, L. V. Piskach, Y. E. Romanyuk, I. D. Olekseyuk, V. I. Zaremba, and V. I. Pekhnyo, J. Alloys Compd. 397, 85 (2005).
5. Teodor K. Todorov, Kathleen B. Reuter, and David B. Mitzi, Adv. Mater. 22, E156 (2010).
6. J. Seol, S. Lee, J. Lee, H. Nam, and K. Kim, Sol. Energy Mater. Sol. Cells 75, 155 (2003).
7. Kejia Wang, Byungha Shin, Kathleen B. Reuter, Teodor Todorov, David B. Mitzi, and Supratik Guha, Appl. Phys. Lett. 98, 051912 (2011).
8. Levent Gütay, Alex Redinger, Rabie Djemour, and Susanne Siebentritt, Appl. Phys. Lett. 100, 102113 (2012).
9. A. Weber, S. Schmidt, D. Abou-Ras, P. Schubert-Bischoff, I. Denks, R. Mainz, and H. W. Schock, Appl. Phys. Lett. 95, 041904 (2009).
10. Alexey Shavel, Jordi Arbiol, and Andreu Cabot, J. Am. Chem. Soc. 132, 4514 (2010).
11. Chet Steinhagen, Matthew G. Panthani, Vahid Akhavan, Brian Goodfellow, Bonil Koo, and Brian A. Korgel, J. Am. Chem. Soc. 131, 12554 (2009).
12. K. Doverspike, K. Dwight, and A. Wold, Chem. Mater. 2, 194 (1990).
13. S. Levcenco, D. O. Dumcenco, Y. P. Wang, J.D. Wu, Y.S. Huang, E. Arushanov, V. Tezlevan, K.K. Tiong, Optical Materials 34, 1072 (2012).
14. S. Schorr, H. J. Hoebler, and M. Tovar, Eur. J. Mineral. 19, 65 (2007).
15. Clas Persson, J. Appl. Phys. 107, 053710 (2010).
16. Shiyou Chen, Aron Walsh, Ye Luo, Ji-Hui Yang, X. G. Gong, and Su-Huai Wei, Phys. Rev. B 82, 195203 (2010).
17. R. Magri, S. H. Wei, and A. Zunger, Phys. Rev. B 42, 11388 (1990).
18. Shiyou Chen, X. G. Gong, Aron Walsh, and Su-Huai Wei, Appl. Phys. Lett. 94, 041903 (2009).
19. J. E. Bernard, L. G. Ferreira, S. H. Wei, and A. Zunger, Phys. Rev. B 38, 6338 (1988).
20. Shiyou Chen, X. G. Gong, Aron Walsh, and Su-Huai Wei, Phys. Rev. B 79, 165211 (2009).
21. Joachim Paier, Ryoji Asahi, Akihiro Nagoya, and Georg Kresse, Phys. Rev. B 79, 115126 (2009).
22. Silvana Botti, David Kammerlander, and Miguel A. L. Marques, Appl. Phys. Lett. 98, 241915 (2011).
23. G. Kresse and J. Furthmueller, Comput. Mater. Sci. 6, 15 (1996).
24. J. Heyd, G. E. Scuseria, and M. Ernzerhof, J. Chem. Phys. 118, 8207 (2003).
25. G. Kresse and D. Joubert, Phys. Rev. B 59, 1758 (1999).
26. H. J. Monkhorst and J. D. Pack, Phys. Rev. B 13, 5188 (1976).
27. S.-H. Wei and A. Zunger, Phys. Rev. B 49, 14337 (1994).
28. Clas Persson, Appl. Phys. Lett. 93, 072106 (2008).

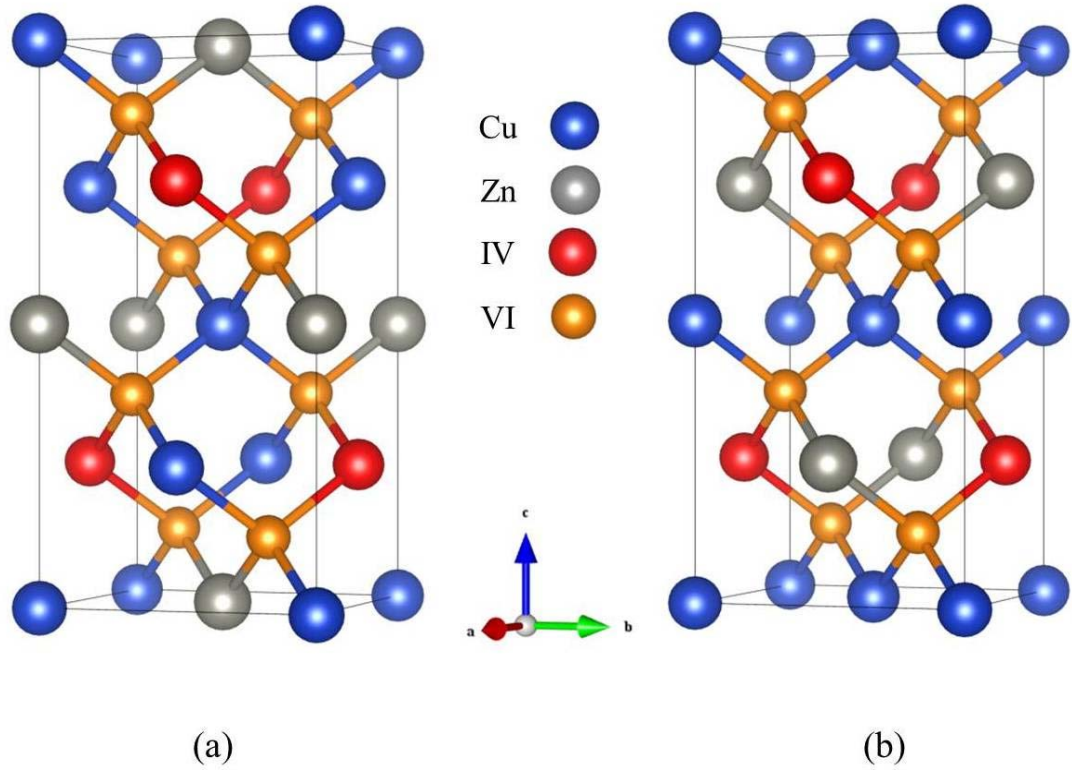

(a) (b)

Figure 1. Conventional crystal cells for (a) the kesterite (KS), and (b) the stannite (ST) structures for $Cu_2Zn-IV-VI_4$ (IV = Si, Ge, Sn and VI = S, Se).

Table 1. The energies, lattice constants, tetragonal distortion parameters, $\eta = c/2a$, and band gaps of zinc-blend-derived $Cu_2Zn-IV-VI_4$. The energies are relative to those of the KS structures and the unit is meV/atom. The unit of lattice constants is Å, and the value of band gap is scaled by 1 eV. We also present the calculated crystal-field splitting $\Delta_{cf}$ and spin-orbit splitting $\Delta_{so}$ and the values are scaled by 1 eV. Experimental data[1-4] are listed for comparison, whereas '-' means that no experimental data is currently available.

|  | $Cu_2ZnSnS_4$ | | | $Cu_2ZnSnSe_4$ | | |
|---|---|---|---|---|---|---|
|  | KS | ST | Exp. | KS | ST | Exp. |
| Energy | 0 | 3.20 | - | 0 | 4.08 | - |
| a | 5.450 | 5.429 | 5.427 | 5.735 | 5.712 | 5.693 |
| η | 0.997 | 1.007 | 1.002 | 0.996 | 1.006 | 0.995 |
| Band gap | 1.48 | 1.30 | 1.44~1.51 | 0.82 | 0.64 | 0.8~1.0 |
| $\Delta_{cf}$ | -0.065 | 0.16 | - | -0.048 | 0.09 | - |
| $\Delta_{so}$ | -0.021 | -0.026 | - | 0.25 | 0.23 | - |
|  | $Cu_2ZnGeS_4$ | | | $Cu_2ZnGeSe_4$ | | |
|  | KS | ST | Exp. | KS | ST | Exp. |
| Energy | 0 | 5.81 | - | 0 | 6.46 | - |
| a | 5.350 | 5.305 | 5.341 | 5.640 | 5.602 | 5.606 |
| η | 0.986 | 1.009 | 0.984 | 0.987 | 1.006 | 0.985 |
| Band gap | 2.07 | 1.77 | 1.94~2.25 | 1.12 | 0.83 | 1.29~1.52 |
| $\Delta_{cf}$ | -0.088 | 0.31 | - | -0.078 | 0.18 | - |
| $\Delta_{so}$ | -0.018 | -0.015 | - | 0.25 | 0.24 | - |
|  | $Cu_2ZnSiS_4$ | | | $Cu_2ZnSiSe_4$ | | |
|  | KS | ST | Exp. | KS | ST | Exp. |
| Energy | 0 | 6.94 | - | 0 | 7.58 | - |
| a | 5.309 | 5.254 | - | 5.607 | 5.542 | - |
| η | 0.978 | 1.006 | - | 0.977 | 1.010 | - |
| Band gap | 3.02 | 2.55 | - | 1.92 | 1.53 | - |
| $\Delta_{cf}$ | -0.063 | 0.47 | - | -0.069 | 0.31 | - |
| $\Delta_{so}$ | -0.027 | -0.033 | - | 0.23 | 0.21 | - |

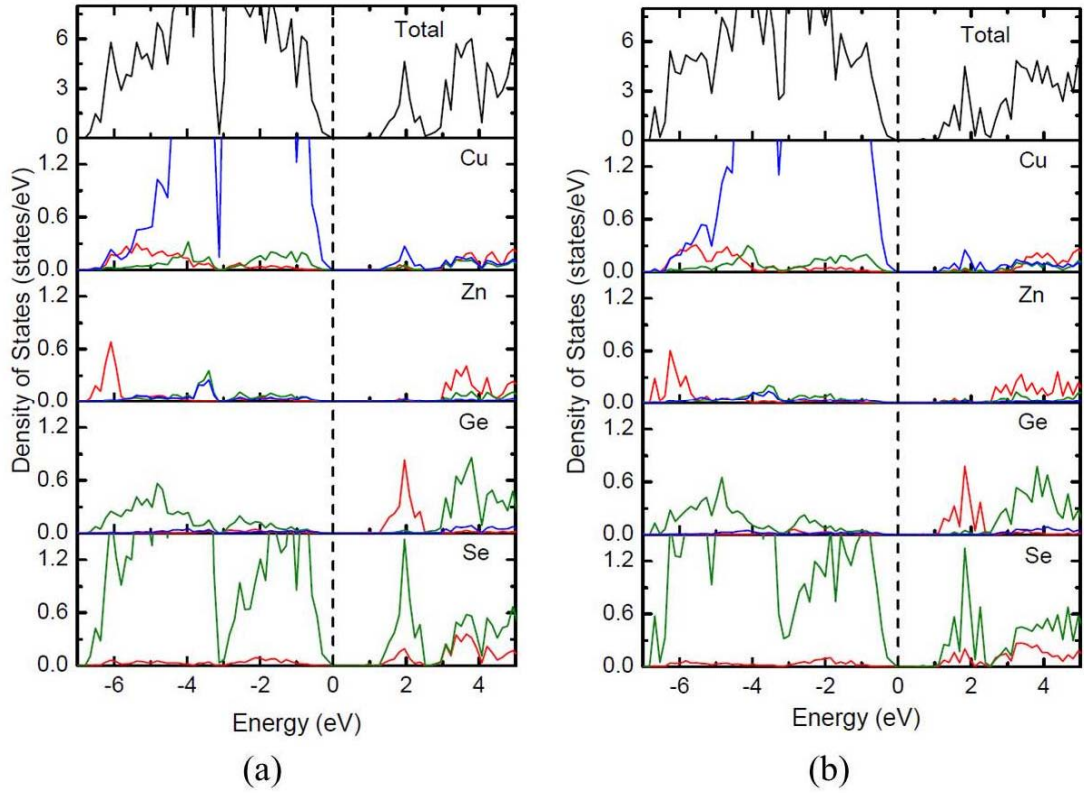

Figure 2. The calculated total and partial DOS of $Cu_2ZnGeSe_4$ by the HSE06 functional: (a) for the KS structure and (b) for the ST structure. The red lines represent the s states, the green lines represent the p states and the blue lines represent the d states. The energies are relative to the valence band maximum (VBM). Note that the y-axes of the first panel and the rest ones are scaled by different factors.

Table 2. The $\Gamma$ point electron effective masses ($m_c$) and hole effective masses ($m_n$ for n = v1, v2, and v3, where v1 is the topmost valence band ) calculated from the band-energy dispersions. The longitudinal masses $m^{\parallel}$ are determined from the dispersions along the (001) direction (that is, the c-axis of the crystal cell shown in Figure 1), while the transverse masses $m^{\perp}$ are derived from the dispersions in both the (110) and (100) directions. The values are scaled by the rest mass of electron, $m_e$.

|  | $Cu_2ZnSnS_4$ | | $Cu_2ZnGeS_4$ | | $Cu_2ZnSiS_4$ | |
|---|---|---|---|---|---|---|
|  | Kesterite | Stannite | Kesterite | Stannite | Kesterite | Stannite |
| $m_{c1}^{\parallel}$ | 0.18 | 0.18 | 0.21 | 0.22 | 0.26 | 0.29 |
| $m_{c1}^{\perp}$ | 0.19 | 0.19 | 0.22 | 0.21 | 0.24 | 0.22 |
| $m_{v1}^{\parallel}$ | 0.22 | 0.74 | 0.27 | 0.67 | 0.40 | 0.71 |
| $m_{v1}^{\perp}$ | 0.74 | 0.34 | 0.91 | 0.41 | 2.45 | 0.55 |
| $m_{v2}^{\parallel}$ | 0.65 | 0.69 | 0.65 | 0.63 | 0.67 | 0.66 |
| $m_{v2}^{\perp}$ | 0.34 | 0.32 | 0.37 | 0.38 | 0.47 | 0.52 |
| $m_{v3}^{\parallel}$ | 0.66 | 0.20 | 0.63 | 0.25 | 0.67 | 0.38 |
| $m_{v3}^{\perp}$ | 0.30 | 0.69 | 0.34 | 0.83 | 0.44 | 1.52 |
|  | $Cu_2ZnSnSe_4$ | | $Cu_2ZnGeSe_4$ | | $Cu_2ZnSiSe_4$ | |
|  | Kesterite | Stannite | Kesterite | Stannite | Kesterite | Stannite |
| $m_{c1}^{\parallel}$ | 0.10 | 0.09 | 0.11 | 0.11 | 0.15 | 0.16 |
| $m_{c1}^{\perp}$ | 0.11 | 0.10 | 0.12 | 0.11 | 0.15 | 0.14 |
| $m_{v1}^{\parallel}$ | 0.12 | 0.55 | 0.13 | 0.50 | 0.21 | 0.52 |
| $m_{v1}^{\perp}$ | 0.32 | 0.14 | 0.38 | 0.17 | 0.58 | 0.27 |
| $m_{v2}^{\parallel}$ | 0.53 | 0.13 | 0.51 | 0.20 | 0.54 | 0.39 |
| $m_{v2}^{\perp}$ | 0.16 | 0.23 | 0.18 | 0.23 | 0.26 | 0.32 |
| $m_{v3}^{\parallel}$ | 0.34 | 0.18 | 0.39 | 0.16 | 0.47 | 0.21 |
| $m_{v3}^{\perp}$ | 0.26 | 0.35 | 0.29 | 0.48 | 0.39 | 0.81 |

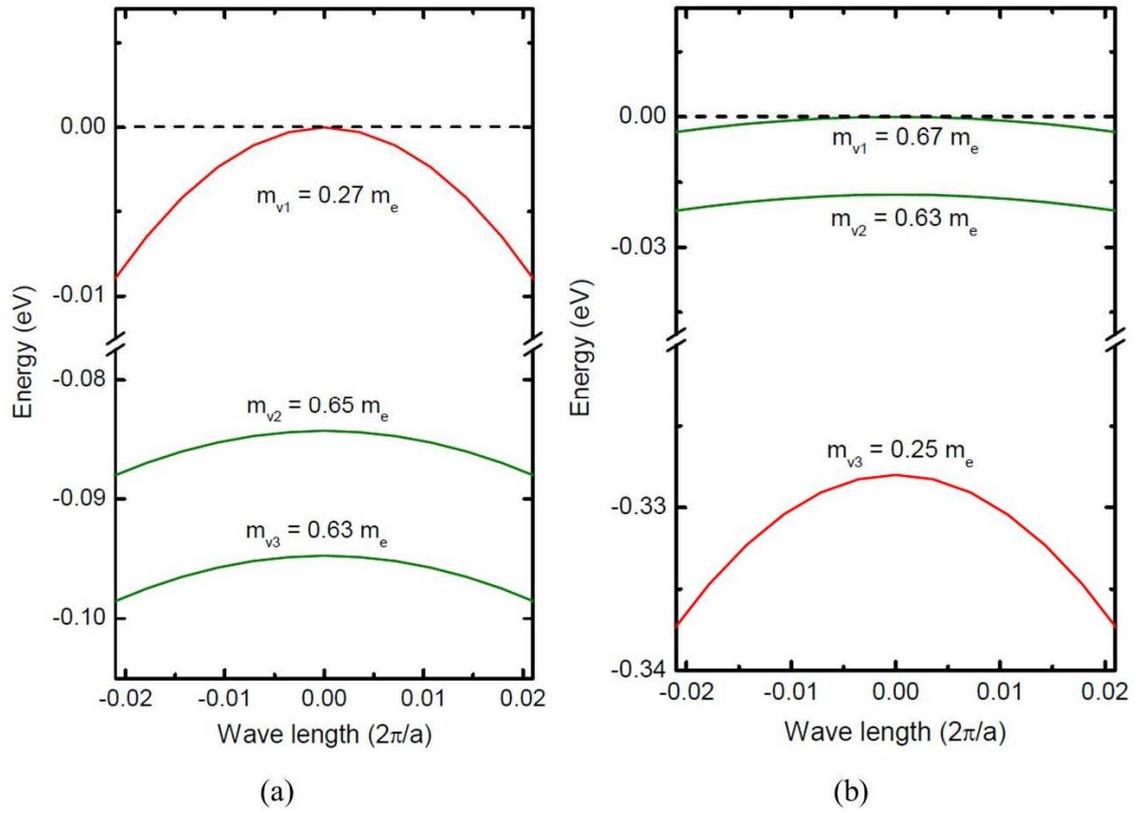

Figure 3. The topmost 3 valence bands along the c-axis of the crystal cell in the immediate neighborhood of $\Gamma$ point: (a) for the KS CZGS, and (b) for the ST CZGS. The VBMs are set to zero.

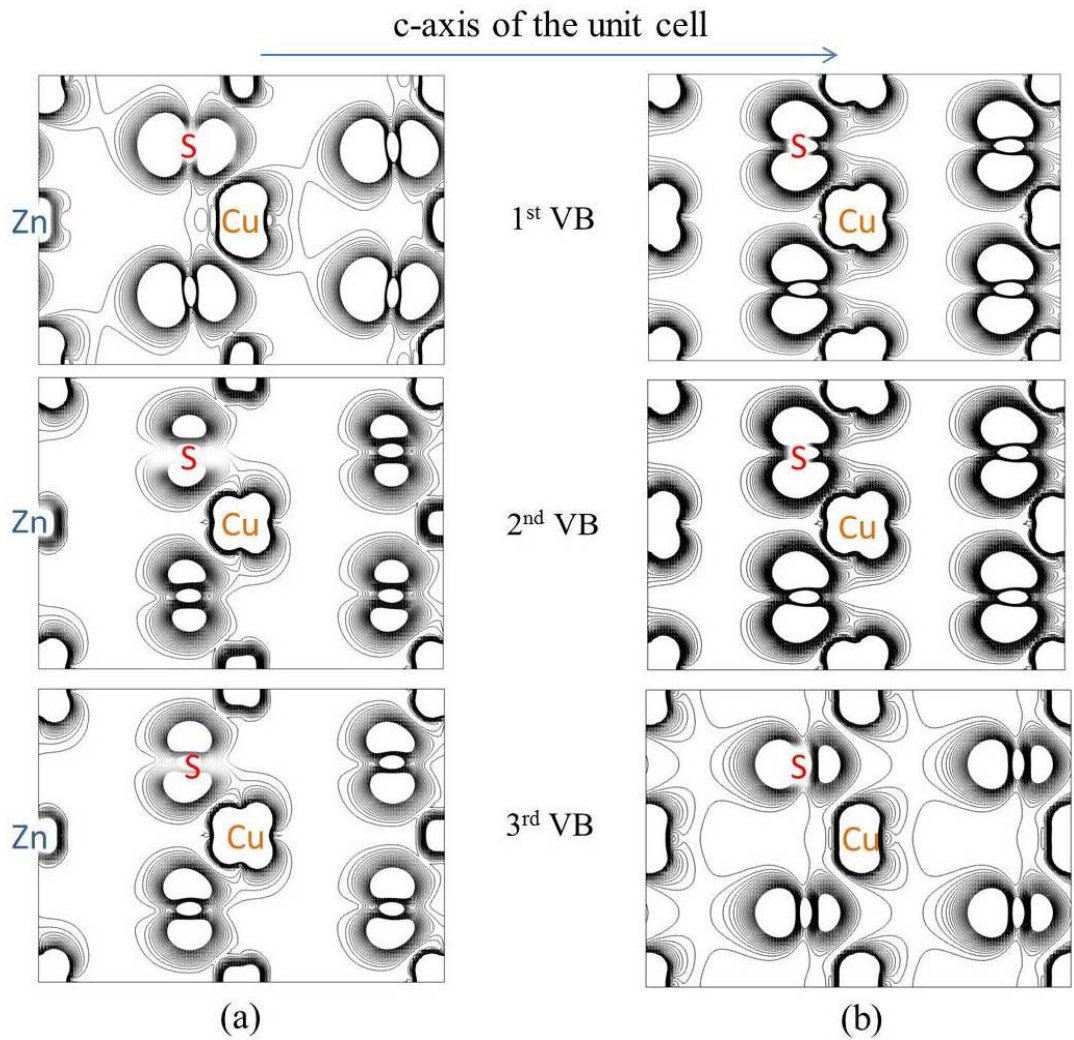

Figure 4. Partial charge density of the topmost 3 VBs at the Γ point in: (a) the Cu-S-Zn plane for the KS CZGS, and (b) the Cu-S plane for the ST CZGS. The direction of the c-axis of the crystal cell is marked by the blue arrow.

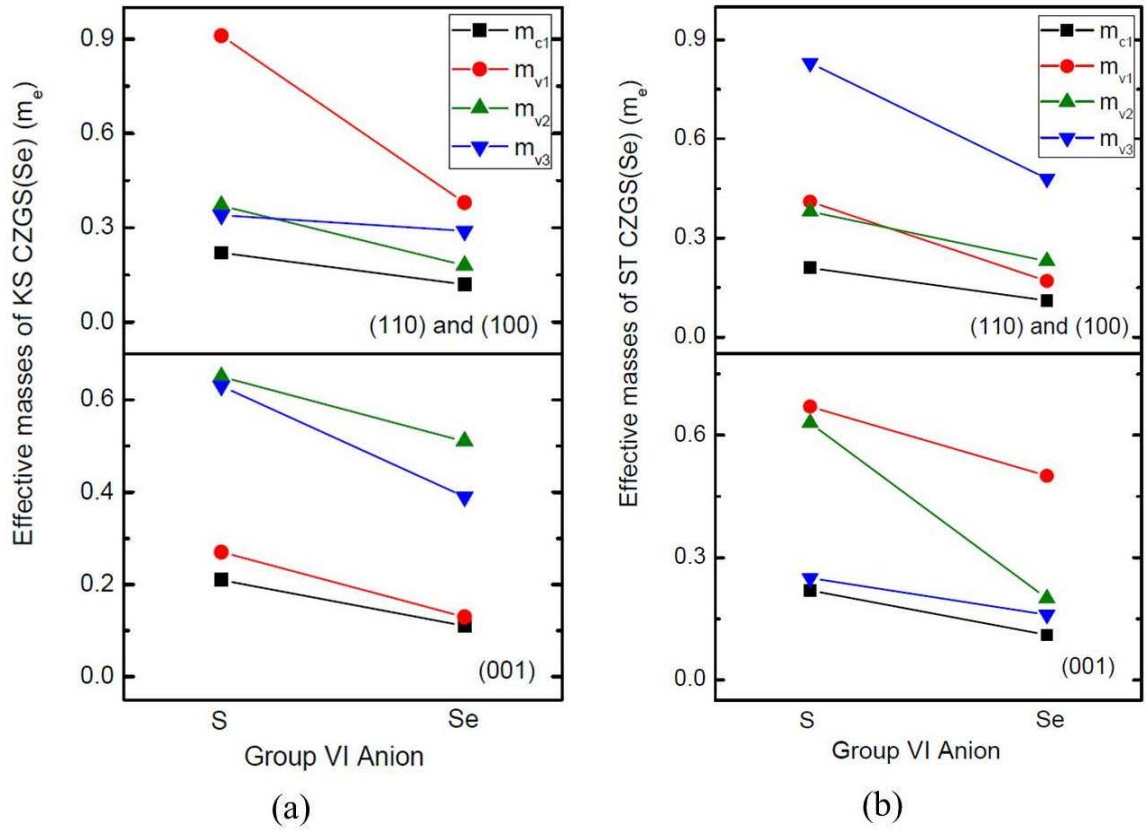

Figure 5. The effective masses of $Cu_2ZnGe - VI_4$ (VI = S, Se) as a function of the group VI anions: (a) for the KS structure, and (b) for the ST structure. The (100) and (110) in the upper panels stands for the direction parallel to the a-axis and the direction parallel to the the diagonal of the square in the ab-plane of the crystal cell shown in Figure 1, respectively, while the (001) in the lower panels stands for the direction parallel to the c-axis of the crystal. The lines are drawn to guide the eye.

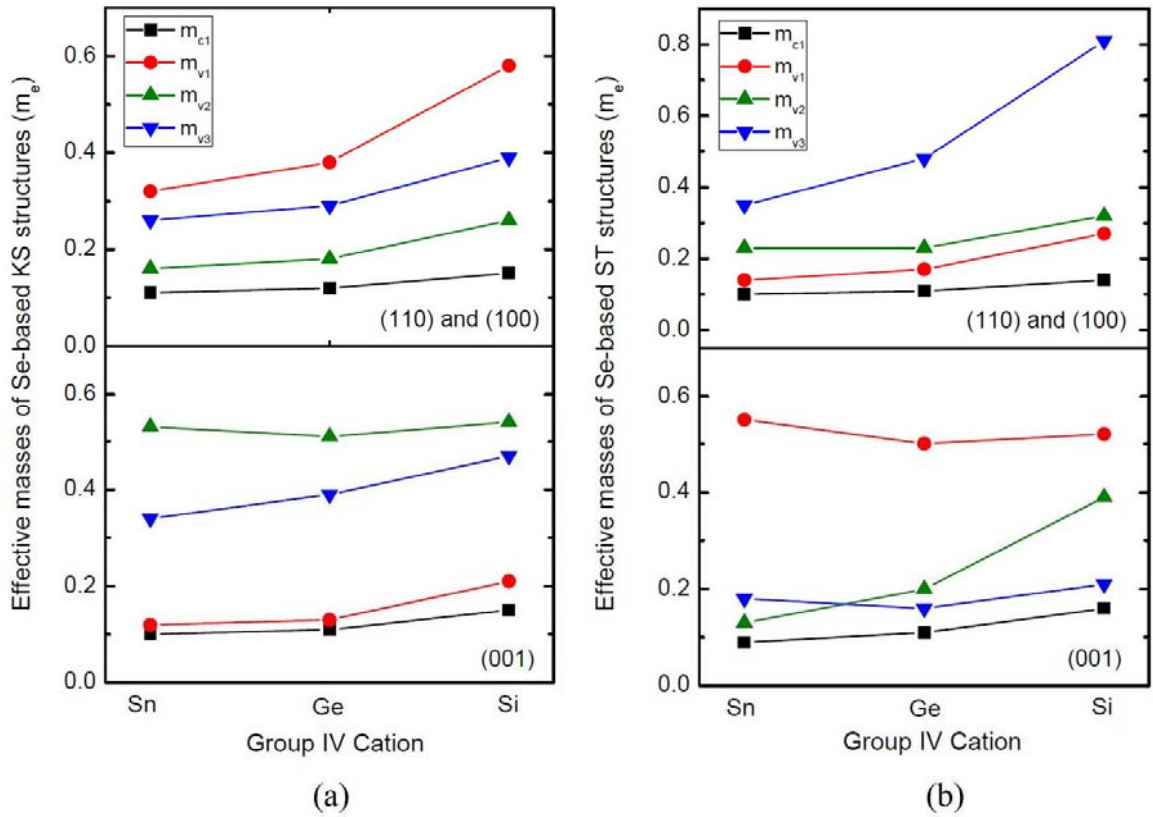

Figure 6. The effective masses of Se-based compounds as a function of the group IV cation: (a) for the KS structure and (b) for the ST structure. The (100) and (110) in the upper panels stands for the direction parallel to the a-axis and the direction parallel to the diagonal of the square in the ab-plane of the crystal cell shown in Figure 1, respectively, while the (001) in the lower panels stands for the direction parallel to the c-axis of the crystal. The lines are drawn to guide the eye.